\documentclass[epj,twocolumn]{webofc}%
\usepackage{graphics}
\usepackage[varg]{txfonts} % Times fonts

\woctitle{ISVHECRI 2018}
\begin{document}
\title{Angular and Abundance Distribution of High-energy Gamma Rays and \\ 
Neutrons Simulated by GEANT4 Code for Solar Flares
}

\author{\firstname{K.} \lastname{Kamiya}\inst{1} \and 
\firstname{K.} \lastname{Koga}\inst{1} \and 
\firstname{H.} \lastname{Matsumoto}\inst{1} \and 
\firstname{Y.} \lastname{Muraki}\inst{2}\fnsep\thanks{\email{muraki@isee.nagoya-u.ac.jp}} \and 
\firstname{S.} \lastname{Shibata}\inst{3}
}
 
\institute{Tsukuba Space Center, JAXA, Tsukuba, 305-8505, Japan \and 
ISEE, Nagoya University, Nagoya, 464-8601, Japan \and
Engineering Science Laboratory, Chubu University, Kasugai 487-0027, Japan
}

\abstract{In the solar flare observed on June 3, 2012, high energy gamma-rays and neutrons were observed.  The event includes  a remarkable feature
of a high neutron/gamma-ratio in  the secondary particles.  We have examined  whether this high n/$\gamma$-ratio can be explained by  simulation.  
As a result of  simulations using the GEANT4 program, the high n/$\gamma$-ratio  may be reproduced for the case that helium and other heavy ions were dominantly accelerated in the flare.
}

\maketitle
\section{Introduction - why we study the sun?} 
\label{intro}

One of the purposes of cosmic-ray research may be found in a research subject to elucidate how and where cosmic rays are accelerated. It has been well-known for a long time that particles are accelerated at the Sun and then arrive at Earth. Particles are also accelerated  in many other places in our universe.  

The high energy particles from the Sun are called Solar Energetic Particles (SEPs) or Solar Cosmic Rays (SCRs). According to Ground Level Enhancement measurements, ions are known to be accelerated up to 10 GeV. How are such particles accelerated to such high energies? This may be a big riddle in astrophysics. A great merit of studying the Sun may be found in a point that we can simultaneously measure solar phenomena with several detectors at different wavelengths. Once upon a time, Professor Sachio Hayakawa, one of the fathers of gamma-ray astronomy, explained  in a lecture (1965) “why astronomy is known as astrophysics”. He stated that although astronomical phenomena are difficult to observe repeatedly, similar phenomena can be tested in our laboratory. This is the reason why astronomy has been involved in the field of physics.

His philosophy is important and is reflected even today in multi-messenger astronomy. The same scientific trend is also seen in solar physics. We are trying to confirm the acceleration process of particles with the use of various instruments located not only at ground level but also in space. Such simulations could possibly be a powerful method of investigating the acceleration process. 

Part of this paper has been presented at the poster session of the conference. Here we report our presentation as the full paper. We have investigated the interaction process of accelerated solar particles in the solar atmosphere by using the GEANT4 simulation code. This paper presents the calculation results.

\section{Impulsive Flares and Gradual Flares} 
\label{Flares}

In this paper, we do not discuss how particles are accelerated on the solar surface. Instead, we present the simulation results of  secondary particles such as gamma rays and neutrons emitted by the collisions that occur when accelerated particles hit the solar atmosphere.

Solar flares are traditionally classified into two categories: the impulsive flare and the gradual flare. The impulsive phase usually lasts about one hour, while the gradual flare continues for about one day. Although several authors have reported on long-duration gamma-ray flares (LDGRFs), their origin 
has yet to be established.  

\begin{figure*}[ht]
\centering
\includegraphics[width=14cm]{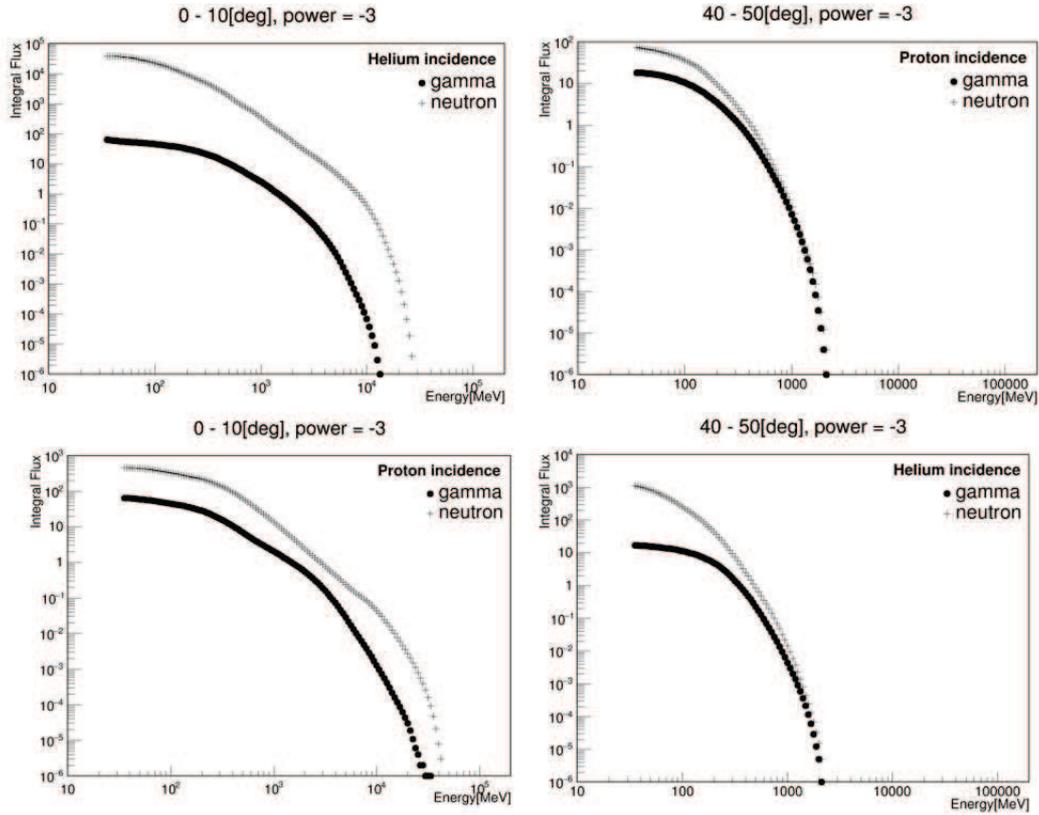}
\caption{Integral energy spectra of neutrons and gamma rays with energy > 100 MeV expected in the very forward region (0-10$^{\circ}$) and intermediate 40-50$^{\circ}$ regions from the initial injection beam. In the case of helium ion entry, a high n/$\gamma$-ratio up to 1,000 is expected.}
\label{fig-1}       % Give a unique label
\end{figure*}

Reames \cite{b1} analyzed several solar flares and concluded that the abundance of helium components and other ions are apparently higher in impulsive flares than in gradual flares. One of the assumptions to explain such a conclusion may be found in the paper by Fisk \cite{b2}. According to his theory, the He$^{3}$ ions and Fe ions in impulsive flares are selectively accelerated by the ion cyclotron resonance process between the waves and the particles under a condition of k$\parallel$v$\parallel$ = $\omega-2 \Omega_{i}$ (a kind of surfing acceleration mechanism).  
In this paper, the expected  ratio of neutrons versus gamma rays (n/$\gamma$-ratio) has been calculated  using the GEANT4 simulation code   (Fig. \ref{fig-1}).

\begin{figure*}[ht]
\centering
\includegraphics[width=5cm]{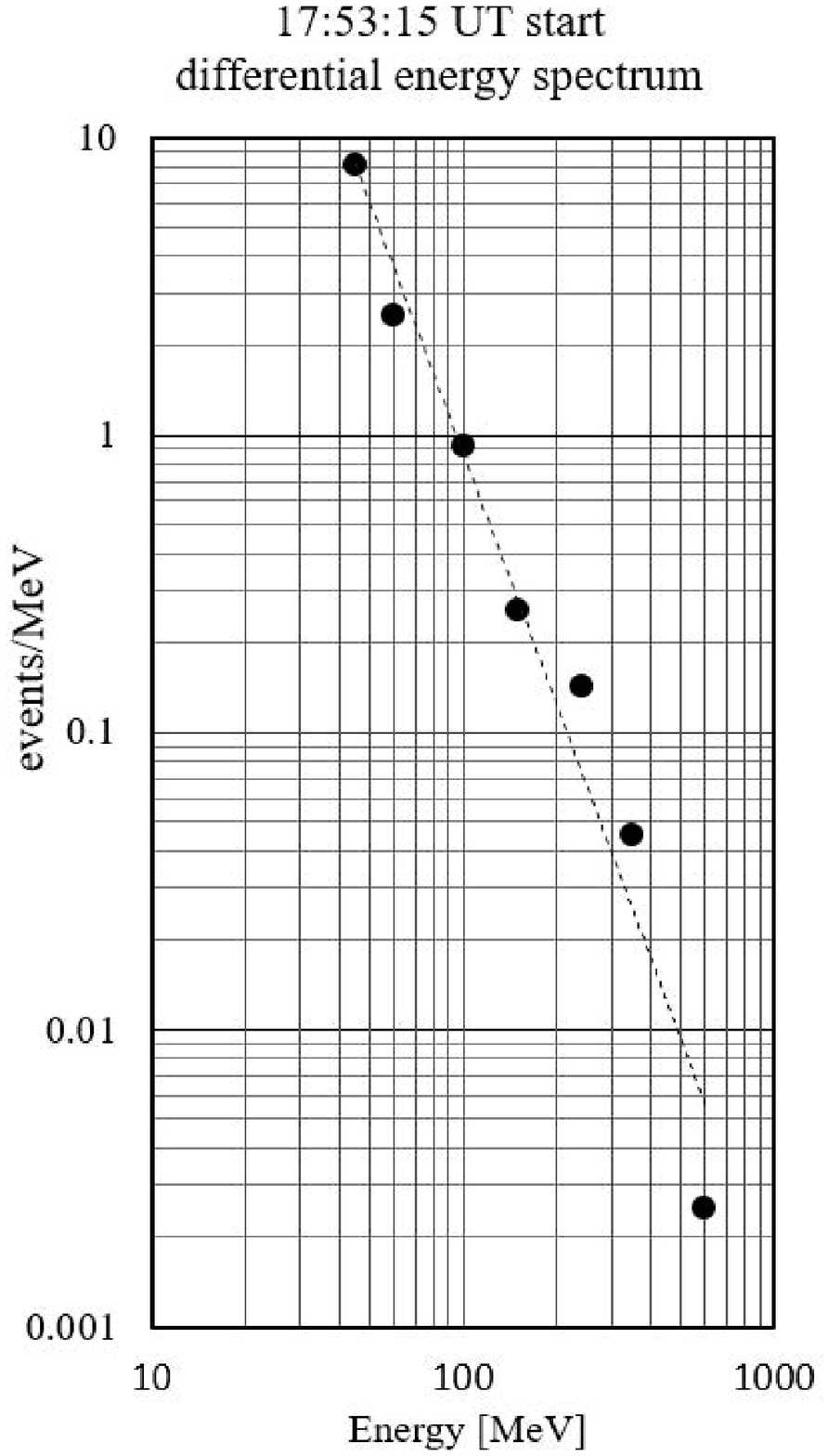}
\includegraphics[width=8cm]{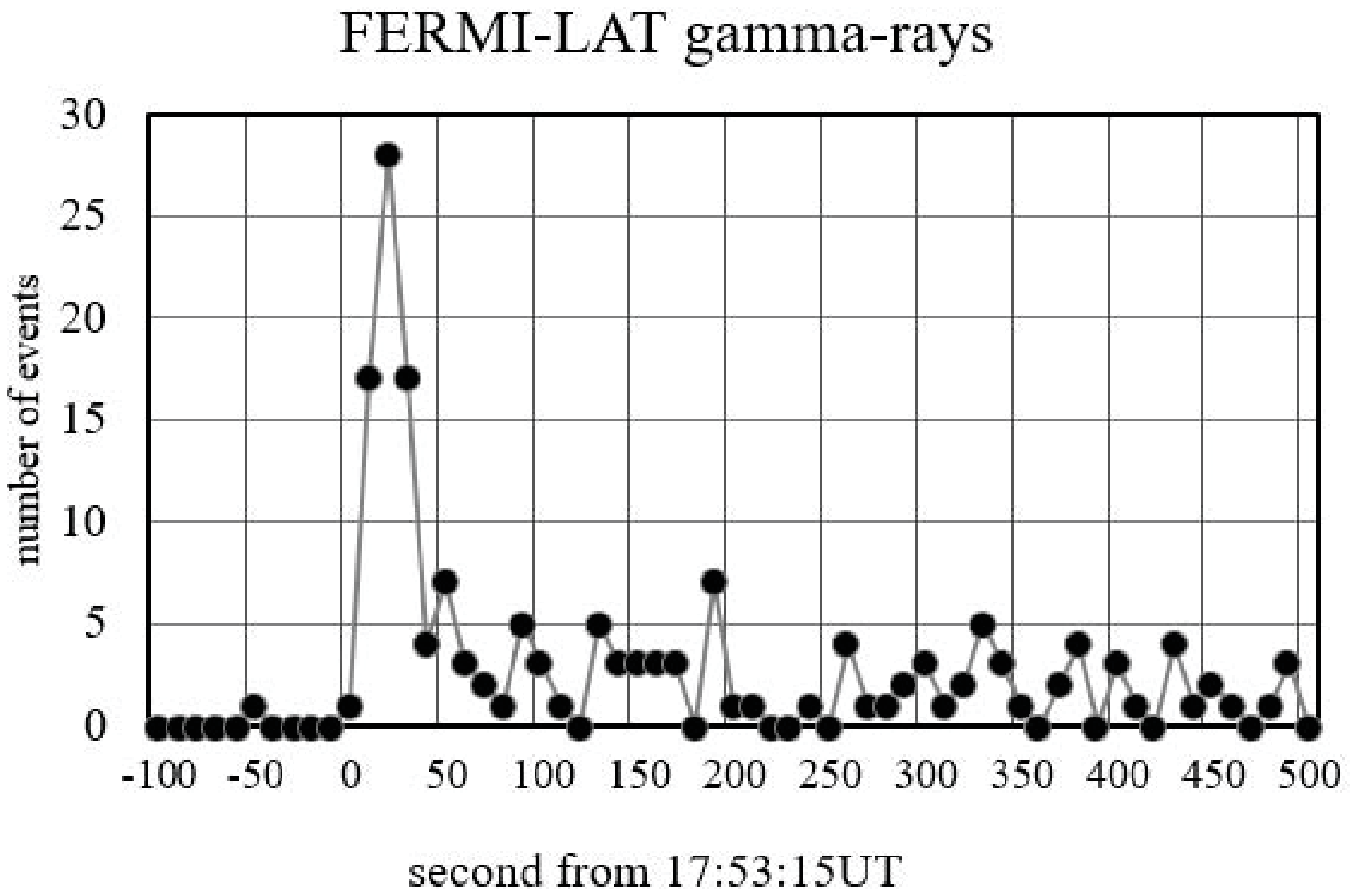}
\caption{(a) Energy spectrum of neutrons observed by SEDA-FIB  \cite{b3}; (b) time profile of gamma rays observed by FERMI-LAT  \cite{b4}. The power index of neutrons is given by (2.8 $\pm$ 0.3).}
\label{fig-2}       % Give a unique label
\end{figure*}

%\newpage
\section{Angular Distribution of Secondary Particles}
\label{Angular}

First, we bombarded one million protons and helium ions vertically to the solar surface without assuming the mirroring effect 
(i.e., pitch angle = 0$^{\circ}$).
 The energy of protons was changed from E$_{p}$ = 100, 200 and 500 MeV, and 1, 5, 10, 50 and 100 GeV.  We have estimated the fluxes of the secondary particles 
for the case of the solar atmospheric depths, the equivalent gram-forces of 10 g/cm$^{2}$.  The atmospheric depth of 10 g/cm$^{2}$ corresponds to z=320  km inside the photosphere.   
The upper parts of Figure \ref{fig-1} show the integral energy spectrum of secondary neutrons and gamma-rays emitted toward the forward direction of 0$^{\circ}$-10$^{\circ}$ and 40$^{\circ}$-50$^{\circ}$  for  proton incidence; Flux(>E$\gamma$) = $\int$f(E$\gamma$)dE$\gamma$ and 
Flux(>E$_{n}$) =  $\int$f(E$_{n}$)dE$_{n}$.  
The lower pictures of Figure \ref{fig-1} represent the same plot, but for  helium incidence.  For  helium ions, similar energies were selected but the collision energy is set 
for the incident energy per nucleon of the helium nucleus. The n/$\gamma$ ratio at E$_{n,\gamma}$~>~100~MeV is predicted as 7.5 for  proton incidence,
 but a higher n/$\gamma$-ratio such as $\sim$1,000 is predicted for  helium incidence.  
This may provide us with a possible explanation of the high n/$\gamma$-ratio caused by the threshold effect, 
arising from the difference between proton and helium incidences.  As shown in Figures \ref{fig-1} and \ref{fig-2}, the observed high n/$\gamma$-ratio may suggest that the dominant acceleration of helium ions possibly occurred in the flare of 3 June 2012.

\begin{figure*}[ht] 
\centering
\includegraphics[width=16cm]{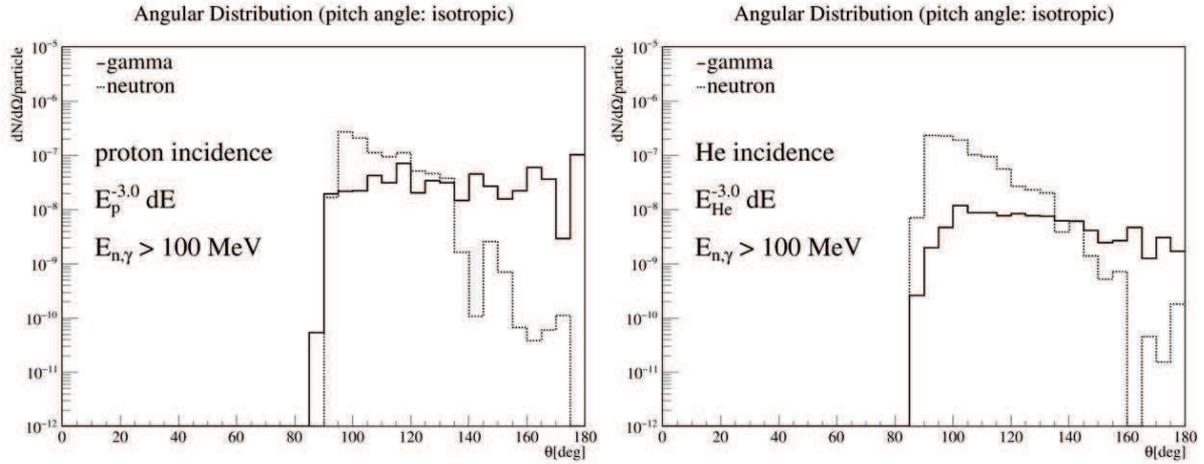}
\caption{Angular distribution of the emitted secondary particles (n, $\gamma$). The initial proton and helium ions are assumed to have the energy spectrum expressed by power index E$^{-3}$. Note that 90$^{\circ}$ corresponds to the direction of the solar center.}
\label{fig-3}       % Give a unique label
\end{figure*}

\section{Comparison with Observed Results} 

In order to compare the simulation results with the observed results, we selected the solar flare event observed on June 3, 2012. This flare showed a distinct feature and condition: 
\begin{enumerate}
\item it only lasted up to one minute, and 
\item both high energy neutrons and gamma rays were detected by the two detectors onboard the ISS  \cite{b3} and the Fermi satellite  \cite{b4}.  
\end{enumerate}

The feature described in (1) is important to obtain the energy spectrum of solar neutrons, as we can fix the production time of neutrons to within one minute. 
The condition described in (2) is important to estimate the abundance of ions in the flare particles.

Figures \ref{fig-2} (a) and (b) show the energy spectrum of solar neutrons and the production time profile of gamma rays with energy higher than 100 MeV, respectively. In Figure \ref{fig-2} (a), we have assumed the instantaneous production time of solar neutrons at 17:53:15 UT. Even if we assume the production time at 17:53:35 UT, the maximum production time of gamma rays, the effect of the start time on the power index of the energy spectrum of neutrons is less than the statistical error ($\pm$ 0.3).

When we obtain the n/$\gamma$-ratio, the value turns out to be (1110 $\pm$ 350). According to the simulation,  the ratio of n/$\gamma$ for proton incidence is expected to be (1.3 $\pm$ 0.5) (Fig. \ref{fig-3} (a)), while that for helium incidence is expected to be (1500 $\pm$ 600) (Fig. \ref{fig-3} (b)). If the helium component involved in the flare increases beyond 36\%, the observed high n/$\gamma$-ratio (1110 $\pm$ 350) could be explained (as the value is obtained by the 1$\sigma$ overlap region between the p and He values).  

\section{Conclusion}

We attempted to explain the 2012.6.3 event based on the simulation results, by focusing on the flux of secondary particles between the high energy neutrons and gamma rays emitted by collisions. From the coordinates of the solar flare (N16$^{\circ}$E33$^{\circ}$), we estimated the direction to Earth as viewed by the flare position. In other words, the direction to Earth is 53$^{\circ}$ above the solar horizon. According to our simulation, a high 
n/$\gamma$-ratio is predicted for the secondary particles emitted in the very forward cone (less than 10$^{\circ}$). 
Therefore, in order to explain the high n/$\gamma$-ratio, the injection angle of the magnetic field must be inclined up to about 50$^{\circ}$ from the vertical direction of the solar surface. The ions trapped in the magnetic field then fall down into the solar atmosphere; some escape  from the solar surface by mirroring, and some may interact with the solar atmosphere and produce  secondary particles such as charged  and neutral pions, and cause the breakup of ion nuclei. 
There may be a chance to theoretically explain the event observed on June 3, 2012, provided that the production process satisfies the above conditions in the simulation. Our analysis is consistent with the early prediction by Reames \cite{b1} that heavy components are dominantly involved in the  particles accelerated in impulsive flares. 

\section*{Acknowledgments}

We wish to thank  Dr. Yasuyuki Tanaka for the reduction of the FERMI-LAT data of June 3, 2012. Y.M. is supported by the Grant-in-Aid of Scientific Research (C) 16k05377 from the Japan Society of the Promotion of Science.


\begin{thebibliography}{99} 

\bibitem{b1} Reames, D.V. (1994) Adv. Space Res., 15(7), 41, and (1995) {\it AIP Conference Proceeding}, 374, High Energy Solar Physics, eds. R. Ramaty, N. Mandzhavidze and X.-M. Hua, 35-45.

\bibitem{b2} Fisk, L.A., ApJ, 224 (1978) 1048.

\bibitem{b3} Koga,K. et al, Solar Physics, 292 (2017) 115.

\bibitem{b4} Atwood, W.B. et al., ApJ, 697 (2009) 1071.

\end{thebibliography}
\end{document}